\documentstyle[aps,floats,prl,twocolumn]{revtex}
\begin{document} 

\title{ Comparative Study of the Insulator-Hall Liquid-Insulator Transitions:
    Composite Boson Picture vs Composite Fermion Picture}
\author{Wenjun Zheng, Yue Yu and Zhaobin Su}
\address{Institute of Theoretical Physics, Chinese Academy of Sciences, Beijing
       100080, P.R.China}
\date{\today}       
\maketitle 

\begin{abstract}
\noindent
  Based on a newly advanced phenomenological understanding of the high-field insulator-
Hall liquid transition in a composite fermion picture,  we extend its composite
boson counterpart to the analysis of the low-field insulator-Hall liquid transition.
We thus achieve a comparative study of these two transitions. In this
way, the similar reflection symmetries in filling factors in both transitions
are understood consistently as due to the symmetry of the gapful excitations 
which dominate $\sigma_{xx}$ across the transitions, and the abrupt change in 
$\sigma_{xy}$ at the transitions. The substantially different characteristic
energy scales involved in these two transitions can be attributed to the differences
in critical filling factors $\nu_c$ and the effective masses. The opposite temperature-dependences of the
critical longitudinal resistivities are also well-understood, which can be traced
to the opposite statistical natures of the composite fermion and the composite
boson. We also give a tentative discussion of the zero-temperature dissipative
conductivity. The above results are supported by a recent experiment
(cond-mat/9708239).        
       
\end{abstract}
\vskip 0.1in
 
\vskip 0.1in
    The study of the magnetic field induced insulator-Hall liquid transitions
has evoked considerable interests in the past decade. In the theoretical aspect, Kivelson, 
Lee and Zhang(KLZ)\cite{KLZ} started from the bosonic Chern-Simons field
theory\cite{SCZ} and gave an elegant derivation which lead to the overall phase 
diagram of the general quantum Hall system with respect to disorder and magnetic field.
A qualitatively identical phase diagram was obtained by Halperin, Lee and Read
\cite{HLR}, from the celebrated composite fermion(CF) theory\cite{Jain}. The well-known
correspondence rule advanced by KLZ established a series of connections between
the plateau-plateau transitions and the insulator-Hall liquid transitions,
which suggests the possibility of super-universality\cite{SU} in the diverse 
quantum phase transitions observed in the quantum Hall system\cite{S1}\cite{S2}.
 Many experiments have been conducted to check these ideas\cite{HI}. 
 
     In a recent experiment\cite{Tsui}, Hilke et al. examined the magnetic field
driven insulator-quantum Hall liquid-insulator transitions of the two dimensional hole
system( 2DHS) in a Ge/SiGe quantum well. With the increase of magnetic field, they found
interesting similarities between the low-field(LF) insulator-Hall liquid transition
and the high-field(HF) Hall liquid-insulator transition, with respect to the transport
 properties. First, the critical longitudinal resistivity at the LF transition
 $\rho^L_c$ and the one at the HF transition $\rho^H_c$ are approximately equal(Fig.1),
 $$\rho^L_c=\rho^H_c\pm 3\%.  $$
 Secondly, there is a reflection symmetry in $\rho_{xx}$ at the LF transition similar to
  the one at the HF transition previously reported by Shahar et al.\cite{Sha}. These
  relations can be fitted by the following equations(Fig.2):
 \begin{equation}
\label{EQ0} 
\rho_{xx}^L(\nu, T)=\rho^L_c\exp[\frac{\Delta
\nu}{\alpha_0^L(T)}], 
\end{equation}  
where $\Delta\nu=\nu-\nu_c^L$, $\alpha_0^L(T)=\alpha^LT+\beta^L$, and $\alpha^L$,
$\beta^L$ are sample-dependent parameters;
\begin{equation}
\label{EQ1} 
\rho_{xx}^H(\nu, T)=\rho^H_c\exp[\frac{-\Delta
\nu}{\alpha_0^H(T)}], 
\end{equation} 
where  $\Delta\nu=\nu-\nu_c^H$, $\alpha_0^H(T)=\alpha^HT+\beta^H$
and $\alpha^H$, $\beta^H$ are sample-dependent parameters;

In spite of the above similarities which suggest similar mechanisms for the 
two transitions, Hilke et al. also pointed out several differences
between the HF transition and the LF transition: First, there are quite different
characteristic energy scales involved in these two transitions, i.e.
$$ \alpha^L/\alpha^H\approx 0.19/0.03\approx 6,$$ 
as the measurement in ref\cite{Tsui} showed(see FIG.2).
Secondly, the temperature-dependences of $\rho^L_c$ and $\rho^H_c$ are opposite,
with the former decreasing with higher $T$ while the later increasing with higher
$T$. Both behaviors show up when $T$ is larger than certain threshold values(FIG.1).
    
     The similar properties of the LF transition and the HF transition seem to
favor the floating up recipe\cite{FU}, where both transitions are attributed to the crossing
of the Fermi level with the lowest extended level. However, the substantially 
different energy scales and the qualitatively opposite $T$-dependences in the critical
resistivities are beyond its predictions. It will be the aim of this paper to
present a consistent phenomenological picture accounting for the above similarities 
and differences.

 In a recent paper\cite{ZYS}, we present a phenomenological picture based on the composite
 fermion theory, in order to understand the reflection symmetry near the 
 transition from a $\nu=1$ quantum Hall liquid to a Hall 
 insulator (the above-mentioned HF transition). In that picture, the seemingly unexpected reflection 
 symmetry in the longitudinal resistivity $\rho_{xx}$ can be understood clearly as due 
 to the symmetry of the gapful excitations which dominate $\sigma_{xx}$ across the transition, 
 and the abrupt change in $\sigma_{xy}$ at the transition. The parameter $\alpha$ in the linear fit of $\nu_0(T)$ 
 in ref\cite{Sha} is also given a simple physical meaning. Based on that
  theory the effective mass can be calculated from $\alpha$, which gives
 a reasonable value of several electron band mass. When taking
 into account the previous network model calculations, the nearly invariant Hall
 resistivity $\rho_{xy}$ across the transition is also well-understood. 

       One can see that the above picture does not directly depend on the statistical nature
of the CFs. That is to say, the CB counterpart of it will produce the same
reflection symmetry in $\rho_{xx}$. Based on this consideration, we attempt to use this CB picture
to describe the LF transition, and then give an explanation of
the properties that are different from the HF transition.
       
Noticing that the critical magnetic field at the HF transition $B_c^H=4.04T$ 
is almost twice the value at the LF transition $ B_c^L=1.975T$, we argue
that the one-flux-quanta bound composite boson will also be a kind of good quasiparticle near
the LF transition, under the condition that the two-flux-quanta bound composite fermion is useful in accounting for
the transport properties near the HF transition. Because the landau level (LL) mixing
is more severe near the LF transition with relatively larger disorder and smaller
LL spacing, the measured critical filling factor $\nu_c^L=1.77$ deviates much
from the ideal lowest LL constrained factor 1. We ignore here the possible contribution
of the particles from the second LL. Therefore, we can suppose that at the LF
transition, there are well-defined CBs in a zero effective magnetic field, while
the CBs will feel the energy gap induced by the effective magnetic field $B^*$ when $\nu$ deviates
from $\nu_c^L$ a little. So we have
\begin{equation}
\label{EQ2}
\sigma_{xx}(\nu, T) \propto \sigma^{CB}_{xx}(\nu, T) \propto\exp(-\omega_c^*/k_B T),   
\end{equation}
where $\omega_c^*=\frac{\hbar eB^*}{m^*_{CB}}  \propto |\nu-\nu_c^L|$, $\hbar$ is set to unit,
and the first relation is derived from the transformation between electrons and
CBs given by ref\cite{KLZ}.

In spite of the above continuity and symmetry around $\nu_c^L$, there should be however a sharp change in
the Hall conductivity $\sigma_{xy}$ across the LF transition point. For $\nu<\nu_c^L$
or the quantum Hall liquid phase, we have $\sigma_{xy}=e^2/h~(T\rightarrow 0)$; while for $\nu>\nu_c^L$ or
the insulator phase, we get $ \sigma_{xy}\rightarrow 0~(T\rightarrow 0)$. This is also consistent with
the well-known "floating up" recipe\cite{FU}, where the QHL-Insulator transition occurs
at the crossing of the Fermi level with the lowest extended state and 
$\sigma_{xy}$ is determined by the number of extended states below the Fermi level.

   With the above results of conductivity,  We can obtain the 
resistivity tensor by conducting an inversion of the conductivity tensor, that is
\begin{eqnarray}
\rho_{xx}=\frac{\sigma_{xx}}{\sigma_{xx}^2+\sigma_{xy}^2}.
\end{eqnarray}
When $\Delta \nu <0$, 
\begin{eqnarray}
\rho_{xx}&\approx& \frac{\sigma_{xx}}{\sigma_{xy}^2}  \\ \nonumber
         &\propto& \sigma_{xx} \\ \nonumber
         &\propto& \exp(-\omega_c^*/k_B T).
\end{eqnarray}
When $\Delta \nu >0$,
\begin{eqnarray}
\rho_{xx}&\approx& \sigma_{xx}^{-1} \\   \nonumber
         &\propto& \exp(\omega_c^*/k_B T).
\end{eqnarray}
Combining the above results and the relation $\omega_c^*\propto |\Delta\nu|$,
 we can easily identify the reflection symmetry,
$$ \rho_{xx}^L(\nu, T)=\rho^L_c\exp[\frac{\Delta\nu}{\alpha^L T}],   $$
where 
\begin{equation}
\label{EQ3}
\alpha^L=\frac{2\pi k_B m^*_{CB} \nu_c^{L2}}{h^2 n},
\end{equation}
 $n$ is the density of the 2DHS. After taking into account the corresponding 
 result for the HF transition ( the case of CFs), we have
$$ \frac{\alpha^L}{\alpha^H}=\frac{\nu_c^{L2}}{\nu_c^{H2}}\frac{m^*_{CB}}{m^*_{CF}}.$$
Since $\nu^L_c\approx 2\nu^H_c$ contributes a factor 4 in the ratio $\alpha^L/\alpha^H\approx 6$,
 we can attribute the substantially different energy scales between the LF transition
 and the HF transition as mainly due to the variant $\nu_c$. Besides, we also 
 expect $\frac{m^*_{CB}}{m^*_{CF}}$ to be larger than 1, because of the different
 extents of disorder for CFs and CBs( see below). In this respect
 the above theory is consistent with  the experimental result of the ratio
 $\alpha^L/\alpha^H\approx 6.$

In another way, one can use the data 
from ref\cite{Tsui} to estimate the effective masses $m^*_{CB}$ and $m^*_{CF}$. Substitute 
$n=0.87\times 10^{11} cm^{-2}$, $\alpha^H=0.03 K^{-1}$ and $\alpha^L=0.19 K^{-1}$ 
into eq (\ref{EQ3}) and its counterpart for CFs in the HF transition, 
one can get,
$$ m^*_{CB}\approx 9m_b ,~~~~~~m^*_{CF}\approx 6m_b,   $$
which give reasonable values of several band mass for $m_b=0.1m_e$. This fact gives support to
our usage of the CB(CF) picture in the LF(HF) transition.

     Then let us turn to the discussion of the critical longitudinal resistivity
$\rho_c^H$ and $\rho_c^L$. At the critical point of the HF(LF) transition, 
the effective magnetic field $B^*$ is averaged to zero. To get started, we adopt the 
simplest picture of free CFs (CBs) moving in a random potential. This picture 
is  not as easy as it seems to be, because the disorder is relatively strong.
( From the measurement, $\rho_c\approx 2.2h/e^2$, so $k_F l$ is of the order 
of 1 or less, which has reached the IR limit). Therefore, for the case of CFs, we
can expect the Drude formula to hold at most marginally, which gives:
\begin{eqnarray}
\sigma_c^{CF}&=&ev_F\frac{dn}{dE}(el_{CF})  \\  \nonumber
           &=&ev_F\frac{n}{E_F}(el_{CF})  \\  \nonumber
           &=&ev_F\frac{k_F^2}{4\pi E_F}(el_{CF})  \\   \nonumber
           &\propto& \frac{e^2}{h}(k_Fl_{CF}).
\end{eqnarray}
One can see that $\rho_c^H$, or its inversion $\sigma_c^{CF}$ is uniquely determined 
by a single dimension-less parameter $k_Fl_{CF}$, which measures the extent 
of disorder. We then make a reasonable extension of the above conclusion to the 
case of CBs, with $k_F$ substituted  by the typical wave vector
$k_{CB}$ specific to the CBs and $l_{CF}$ substituted by its CB counterpart $l_{CB}$, 
We note that $k_{CB}$ is much smaller than $k_F$ at a temperature $T<<E_F$ (we avoid 
 applying the Drude formula directly to the CBs, because its wavelength is much
  larger than $l_{CB}$, and the classical picture is no longer valid). If we 
  suppose that the mean free path is almost the same for CFs and CBs, then this
dimensionless parameter for CBs will be much smaller than that for CFs. Therefore the 
 localization effect of disorder is more severe on CBs than on CFs. So we can expect the 
  disorder potential induced effective mass for CBs to be larger than its 
  counterpart for CFs. This is consistent with above calculation of $m_{CB}/m_{CF}\approx 1.5$.

     The above "single parameter" argument can also be applied to the qualitative 
     analysis of the temperature-dependences of the critical resistivities 
     $\rho_c^H$ and $\rho_c^L$. Let us first suppose a similar $T$-dependence in
the mean free paths for CFs and CBs (that is, they decrease as $T$ increases). Then
for the CF case, only CFs near the Fermi surface have contributions to the 
transport properties, with momentum $k_F$ almost independent of $T$. So the 
parameter $k_Fl_{CF}$ will decrease as $T$ increases,
which implies that $\rho_c^H$ will increase as $T$ goes up. In contrast, for 
the case of CBs, an increase in $T$ will excite the CBs from low momentum states to
higher momentum ones, which results in a considerable increase in $k_{CB}$ that
can counteract the decrease in $l_{CB}$\cite{EXP1}, so the overall tendency for the 
parameter $k_{CB}l_{CB}$ will be an increment. Therefore $\rho_c^L$ will increase as $T$ goes up. In this
way, one can see that the opposite $T$-dependences in $\rho_c^H$ and $\rho_c^L$
come from the presence of a Fermi surface in the CFs and the absence of one in 
the CBs, which has its origin from the opposite statistical natures of fermions
 and bosons. Based on this speculation, we suggest doing the same measurement 
on a sample with its density 10 times smaller and in the same range of temperature(0K to 10K). In this case, $E_F$ will be of the same order of $T$, and the 
Fermi surface effect will be weakened considerably. Therefore the different
 $T$-dependences between $\rho_c^H$ and $\rho_c^L$ should disappear.

      Then we comment briefly on the relation between $\rho_c^H$ and $\rho_c^L$ at $T=0$.
The possible universal relation $\rho_c^H(T=0)=\rho_c^L(T=0)$, as suggested by
Hilke at el. can not be understood easily in the present picture, because it is difficult
to give a reliable analytical equation for a strongly disordered, non-interacting 
bosonic system. Numerical methods for model calculations are suitable in this
respect, which will be the focus of our future work.

      We then turn to a tentative discussion of the origin of the zero-temperature
dissipative conductivity reflected from the non-zero $\beta^H$ and  $\beta^L$.
According to our phenomenological picture, near the HF (LF) transition, CFs (CBs)
move under an effective magnetic field $B^*$. In the single particle approximation,
the CFs or CBs reside on the nearly localized quantum Hall states whose spatial
distributions are proportional to $\exp(-(x-X)^2/2 l_{B^*}^2)$ with the center
$X$ distributes almost uniformly across the plane, where $l_{B^*}$ is the
magnetic length corresponding to $B^*$. At the zero temperature limit, we expect
quantum tunnelings between the quantum Hall states to dominate the transport 
properties. Therefore we suggest that the average tunneling probability $p(T=0)$
which is proportional to $\sigma_{xx}(T=0)$, is determined by
$$ p(T=0) \propto  \exp(-d^2/2 l_{B^*}^2)$$      
where $d$ is the average distance between adjacent particles, or $d\approx 1/\sqrt n$.
Then by using the following relations: 
$$ l_{B^*}=\sqrt\frac{\hbar}{eB^*},~~~\nu^*=nh/eB^*,~~~\nu^*=\frac{\nu \nu_c}{|\Delta\nu|} $$
we can easily arrive at
$$  \sigma_{xx}(T=0)\propto \exp(-\pi\frac{|\Delta\nu|}{\nu_c^2})  $$
Therefore we can estimate 
$$ \beta\approx \frac{\nu_c^2}{\pi} $$
One can compare the above result with the experimentally determined 
$\beta^H$ and  $\beta^L$ by substituting $\nu_c=1/2$ and 1 for the HF and LF
transitions respectively. The theoretical values are
$$\beta^H\approx 0.08,~~~ \beta^L\approx 0.3 ,$$
which give a surprisingly good fit with the experimental data (see Fig.2). 
We comment that in Fig.2b the residue value of $\alpha_0$ should be much closer to 0.3
considering the flattening tendency of the dots when approaching $T=0$.
As we believe, the above consistency should be a very strong support to our
seemingly naive understandings based on CF and CB respectively.

       Before closing, let us give the following comments in order. First, let
us comment on the metallic $T$-dependence for $\rho_c^H$, which is described 
here in the framework of free CFs in a random potential without a magnetic field.
According to the conventional belief\cite{Scale}, a two dimensional non-interacting 
system will be localized to an insulator upon the introduction of an infinitesimal 
extent of disorder, so the metallic phase is absent. However, the situation here is quite
different. Because of the gauge fluctuations in the CF system which break the 
time-reversal symmetry, the weak localization effect\cite{WL} that leads to 
the localization is suppressed. Therefore it is still possible for the CF system to demonstrate
 metallic behaviors at the limit $k_Fl_{CF}\approx 1$. Then, let us discuss 
the role of the Coulomb interaction in the CF system. In a low-density disordered 
two dimensional system, Coulomb gap has important consequences in transport
properties(i.e. a hopping conductivity $\approx \exp(-\sqrt{\frac{T_0}{T}})$)\cite{CG}.
But the effect of Coulomb interaction is different for CFs. Since the particles
are confined to the lowest LL, the CF effective mass has its source from interaction 
and disorder. In Read's recipe\cite{Read}, the residue interactions between the CFs
( the almost neutral flux-hole-electron triplet) are reduced considerably. So
the effect of Coulomb interaction is mainly absorbed into the effective mass of
the CFs. To be complete, we do not exclude the interaction induced $\ln(T)$ correction 
to the conductivity\cite{LN}, which is supposed to be important at quite low 
temperature and irrelevant to the experiment here( the $T$-dependence of 
$\rho_c^H$ manifests for $T$ larger than 3K). As for the CB system, we believe
that the above comments are probably also applicable. Finally, we would like to
suggest that both CFs and CBs are good quasiparticles in quantum Hall systems,
and they are not simply equivalent to each other, nor do they exclude each other. They assume domination in different regimes of the phase diagram, with the excitation energy scales close to the minimum. Because of the opposite statistical
natures of CFs and CBs, we can expect many diverse properties to be observed,
which will be the task of the future experimentalists.

       In conclusion, we have extended a newly advanced CF picture of the 
high-field insulator-Hall liquid transition to its CB counterpart, which is 
then applied to the analysis of the low-field insulator-Hall liquid transition.
We thus present a comparative study of these two transitions. In this
way, the similar reflection symmetries in filling factors in both transitions
are understood consistently as due to the symmetry of the gapful excitations 
which dominate $\sigma_{xx}$ across the transitions, and the abrupt change in 
$\sigma_{xy}$ at the transitions. The substantially different characteristic
energy scales involved in these two transitions can be attributed to the differences
in critical filling factors $\nu_c$ and the effective masses. The opposite temperature-dependences of the
critical longitudinal resistivities are also well-understood, which can be traced
to the opposite statistical natures of the composite fermion and the composite
boson.   We also give a tentative discussion of the zero-temperature dissipative
conductivity, and arrive at a good fit with the experiment. 

We acknowlege helpful discussions with Prof. Y.S.Wu.
We are also thankful to Dr. Shahar for his kindly permission to
our use of the figures. This work is partly supported by National Natural Science Foundation of China.


\input epsf
\begin{figure}
\epsfysize=6cm
\caption{Temperature dependence of the resistivities around the low and 
high field transitions. In fig.\ 2 a) the magnetic fields corresponding 
to the central resistivity curves are 3.94, 4.04 and 4.14 T and in 
fig.\ 2 b) they are 2.05, 1.975 and 1.9 T (reprinted from ref[9]).}
\end{figure}


\input epsf
\begin{figure}
\epsfysize=6cm
\caption{$\alpha_0(T)$ on a linear graph as a function of temperature $T$ for the high-field 
transition. The inset shows the low-field transition up to 1.7 K. The 
straight lines are linear fits to the data (reprinted from ref[9])
.}
\end{figure}


\begin{references}
\bibitem{EXP1} To see this quantitatively, we present the following estimation:
Suppose the CB's average kinetic energy is of the order $k_BT$, then we have
$k_{CB}\sim \sqrt T$. It is reasonable to assume $l_{CB}$ to be near its non-zero value at $T=0$, if the temperature is low. So the product $k_{CB}l_{CB}$ is approximately
proportional to $\sqrt T$.
\end{references}
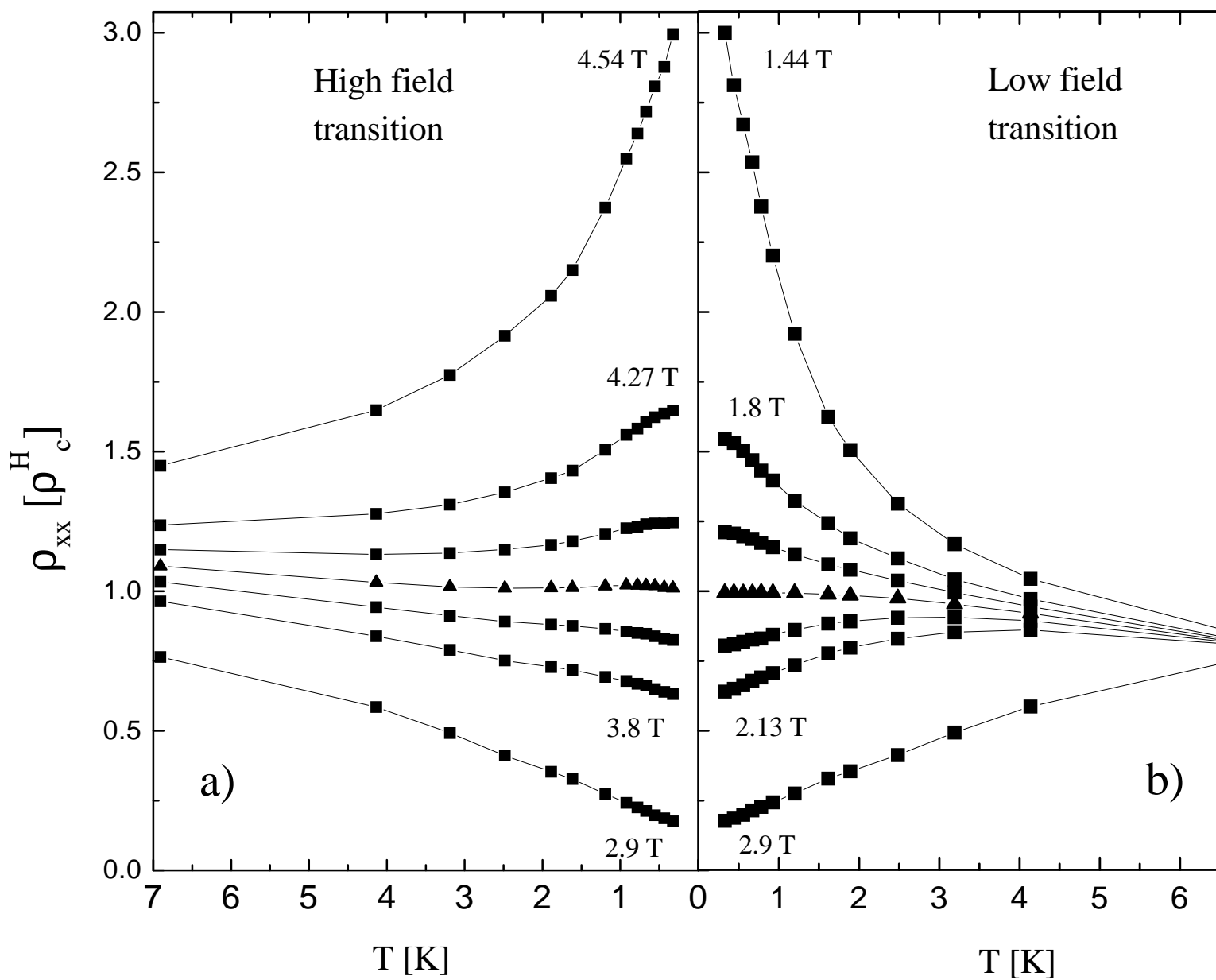
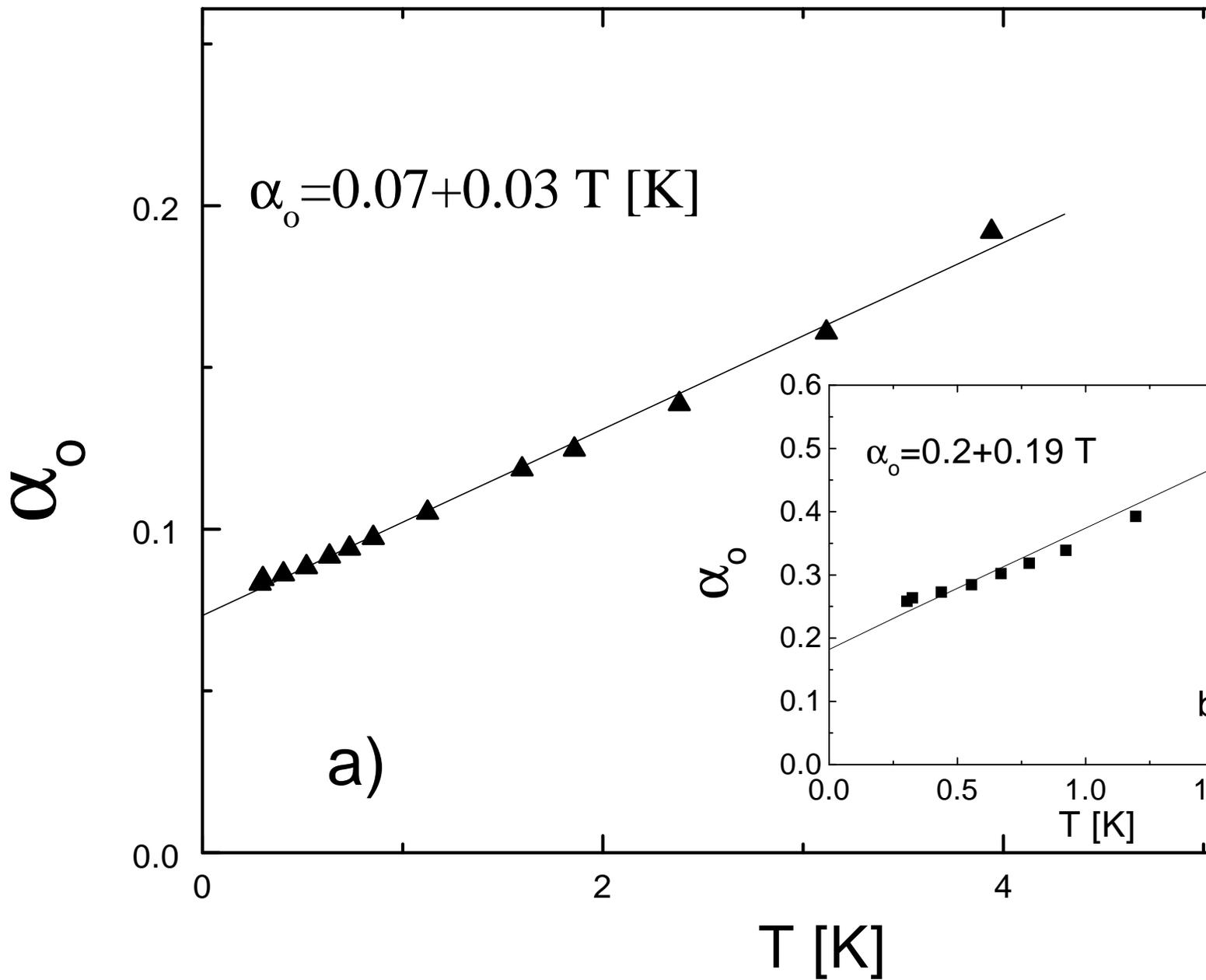
\end{document}